\documentclass{article}
\usepackage[left=1.5cm,right=1.5cm,top=2cm,bottom=2cm]{geometry}

\usepackage{amsmath}
\usepackage{xcolor}
\usepackage{setspace}
\usepackage{graphicx}
\usepackage{authblk}
\usepackage{mathtools}
\usepackage{float}

\usepackage{framed,xcolor}
\definecolor{shadecolor}{gray}{0.95}

\usepackage[hyphens]{url}
\usepackage{hyperref}

\usepackage{multicol}
\usepackage[columnwise]{lineno}

\usepackage{caption}

\usepackage[most]{tcolorbox}

\newif\ifshowboxes
\showboxestrue       

\ifshowboxes
  \newtcolorbox{myshadedbox}{
    colback=gray!15,
    colframe=gray!80!black,
    boxrule=0pt,
    left=1pt,
    right=1pt,
    sharp corners,
    breakable
  }
\else
  
\fi

\definecolor{darkgray}{gray}{0.4}
\usepackage[sorting=none, backend=biber, url=false, date=year, giveninits=true]{biblatex}
\AtEveryBibitem{%
  \clearfield{urldate} 
  \iffieldundef{doi}{}{ 
    \clearfield{url} 
    \clearfield{urldate} 
  }%
}

\DeclareBibliographyDriver{dataset}{%
  \usebibmacro{bibindex}%
  \usebibmacro{begentry}%
  \printnames{author}%
  \newunit\newblock
  \printfield{title}%
  \newunit\newblock
  \printfield{year}%
  \newunit\newblock
  \printfield{publisher}%
  \newunit\newblock
  \printfield{version}%
  \newunit\newblock
  \printfield{doi}%
  \newunit\newblock
  \printfield{url}
  \usebibmacro{finentry}}

\DeclareFieldFormat{urldate}{}
\AtEveryBibitem{\clearfield{month}\clearfield{day}}

\title{A decision-theoretic framework for uncertainty quantification in epidemiological modelling}
\author[1]{Nicholas Steyn}
\author[1]{Freddie Bickford Smith}
\author[1,2]{Cathal Mills}
\author[1]{Vik Shirvaikar}
\author[1,2]{Christl A Donnelly}
\author[3]{Kris V Parag}

\affil[1]{Department of Statistics, University of Oxford}
\affil[2]{Pandemic Science Institute, University of Oxford}
\affil[3]{MRC Centre for Global Infectious Disease Analysis, Imperial College London}

\date{}

\begin{document}

\maketitle


\onehalfspacing

\section*{Abstract}

Estimating, understanding, and communicating uncertainty is fundamental to statistical epidemiology, where model-based estimates regularly inform real-world decisions. However, sources of uncertainty are rarely formalised, and existing classifications are often defined inconsistently. This lack of structure hampers interpretation, model comparison, and targeted data collection. Connecting ideas from machine learning, information theory, experimental design, and health economics, we present a first-principles decision-theoretic framework that defines uncertainty as the expected loss incurred by making an estimate based on incomplete information, arguing that this is a highly useful and practically relevant definition for epidemiology. We show how reasoning about future data leads to a notion of expected uncertainty reduction, which induces formal definitions of \textit{reducible} and \textit{irreducible} uncertainty. We demonstrate our approach using a case study of SARS-CoV-2 wastewater surveillance in Aotearoa New Zealand, estimating the uncertainty reduction if wastewater surveillance were expanded to the full population. We then connect our framework to relevant literature from adjacent fields, showing how it unifies and extends many of these ideas and how it allows these ideas to be applied to a wider range of models. Altogether, our framework provides a foundation for more reliable, consistent, and policy-relevant uncertainty quantification in infectious disease epidemiology.

\clearpage

\begin{multicols}{2}

\section{Introduction}

Statistical epidemiology seeks to learn about infectious disease dynamics by analysing data, often by fitting models \cite{beckerStatisticalStudiesInfectious1999}. These models are used for inference (estimating unobservable quantities such as parameters or latent variables) and/or prediction (estimating hypothetically observable data). Such estimates frequently inform real-world decision-making and public health planning \cite{coriInferenceEpidemicDynamics2024}. In addition to producing estimates, the modeller typically quantifies uncertainty, for example, by presenting confidence intervals, credible intervals, or other uncertainty measures \cite{mccabeCommunicatingUncertaintyEpidemic2021}. 

It is natural to ask where this uncertainty originates, how to quantify it, and how it might be reduced. Beyond its theoretical interest, a robust understanding of uncertainty is essential for determining which aspects of pathogen transmission to model \cite{lloydSensitivityModelBasedEpidemiological2009}, for guiding real-world decision-making \cite{thompsonKeyQuestionsModelling2020}, for planning data collection \cite{caseAdaptingVectorSurveillance2024}, and for communicating results \cite{mccabeCommunicatingUncertaintyEpidemic2021}. Decision-theoretic approaches have been shown to enhance the credibility of using uncertain modelling outputs in these scenarios \cite{millsMetricActionEvaluation2025, bergerRationalPolicymakingPandemic2021}.

Despite its importance, there is little agreement on how to systematically conceptualise and categorise uncertainty in epidemiological models, a lack of consensus also seen in other fields \cite{bickfordsmithRethinkingAleatoricEpistemic2025}. The epidemiological literature on uncertainty is fragmented, with inconsistent definitions, terms, and classifications (for example \cite{dagostinomcgowanQuantifyingUncertaintyMechanistic2021, zelnerAccountingUncertaintyPandemic2021, swallowChallengesEstimationUncertainty2022}). Discussions can be qualitative, lacking the mathematical formalism necessary for reproducibility, comparison between studies, and practical application.

This paper assumes uncertainty arises from making an estimate with incomplete information \cite{fongMartingalePosteriorDistributions2023, kochenderferDecisionMakingUncertainty2015}. Any decision would be fully informed if we had access to ``complete data'', where the precise definition of ``complete data'' depends on the model and problem setting, but uncertainty fundamentally arises from the fact that such data are never available. This leads to an intuitive first-principles decision-theoretic framework for uncertainty quantification which also induces a practical notion of \textit{reducible} and \textit{irreducible} uncertainty.

Our aim is to connect rigorous theoretical ideas with the practical needs of epidemiological modelling by drawing on concepts from machine learning \cite{bickfordsmithRethinkingAleatoricEpistemic2025} and adjacent fields. We adapt and integrate these ideas into a framework for uncertainty quantification in epidemiological modelling by outlining a series of principles while also reflecting on broader conceptualisations. Our framework unifies existing concepts, such as value-of-information (VOI) (from health economics) \cite{jacksonValueInformationSensitivity2019, jacksonValueInformationAnalysis2022, heathValueInformationHealthcare2024},  Bayesian experimental design \cite{ryanReviewModernComputational2016, caseAdaptingVectorSurveillance2024, tanInformationguidedAdaptiveLearning2025, chatzimanolakisOptimalAllocationLimited2020}, information theory \cite{paragQuantifyingInformationNoisy2022, lindleyMeasureInformationProvided1956}, and scoring rules \cite{bosseEvaluatingForecastsScoringutils2024, millsMetricActionEvaluation2025}, in an uncertainty-first form tailored towards epidemiology.

We focus on uncertainty arising within a chosen model and do not attempt to measure the mismatch between the model and the true data-generating process. Statements made about uncertainty here apply to the real world only if the model is correctly specified. However, our framework is a necessary step towards broader uncertainty quantification and we signpost extensions in the discussion.

Section 2 presents our decision-theoretic approach to uncertainty quantification in epidemiology as four principles, motivated by two examples: estimation of infection prevalence and estimation of the instantaneous reproduction number $R_t$, a common measure of epidemic spread \cite{coriNewFrameworkSoftware2013}. Section 3 applies this framework to a real-world model of wastewater surveillance for SARS-CoV-2 in Aotearoa New Zealand. Section 4 discusses how this framework connects, extends, and generalises ideas from adjacent fields. Section 5 concludes with a discussion.

\section{A decision-theoretic framework}

We view model-based uncertainty as arising from missing information \cite{fongMartingalePosteriorDistributions2023, kochenderferDecisionMakingUncertainty2015}. For inference, this missing information is understood as the (possibly infinite) additional data that, if observed, would allow us to precisely determine any identifiable target quantity. For prediction, this view separates a reducible component of uncertainty that vanishes with unlimited data (a notion of parametric uncertainty) from inherent variability in the data-generating process. We describe our approach by stating four principles, motivating them with two conceptual examples.

This missing information perspective is foundational to Bayesian inference (although our framework is not restricted to Bayesian methods), in which a joint distribution is constructed over all modelled random quantities. By conditioning on the observed data, a posterior distribution over unobserved quantities is obtained \cite{robinsConditioningLikelihoodCoherence2000}, leading to the notion that uncertainty ``flows from'' missing data. The modelling of these missing data can be implicit, usually by updating beliefs about model parameters \cite{robertBayesianChoiceDecisionTheoretic2007}, or explicit, by directly modelling the missing data \cite{fongMartingalePosteriorDistributions2023, dawidPresentPositionPotential1984, fortiniExchangeabilityPredictionPredictive2025}.

\begin{figure}[H]
    \centering
    \fbox{\includegraphics[width=\columnwidth]{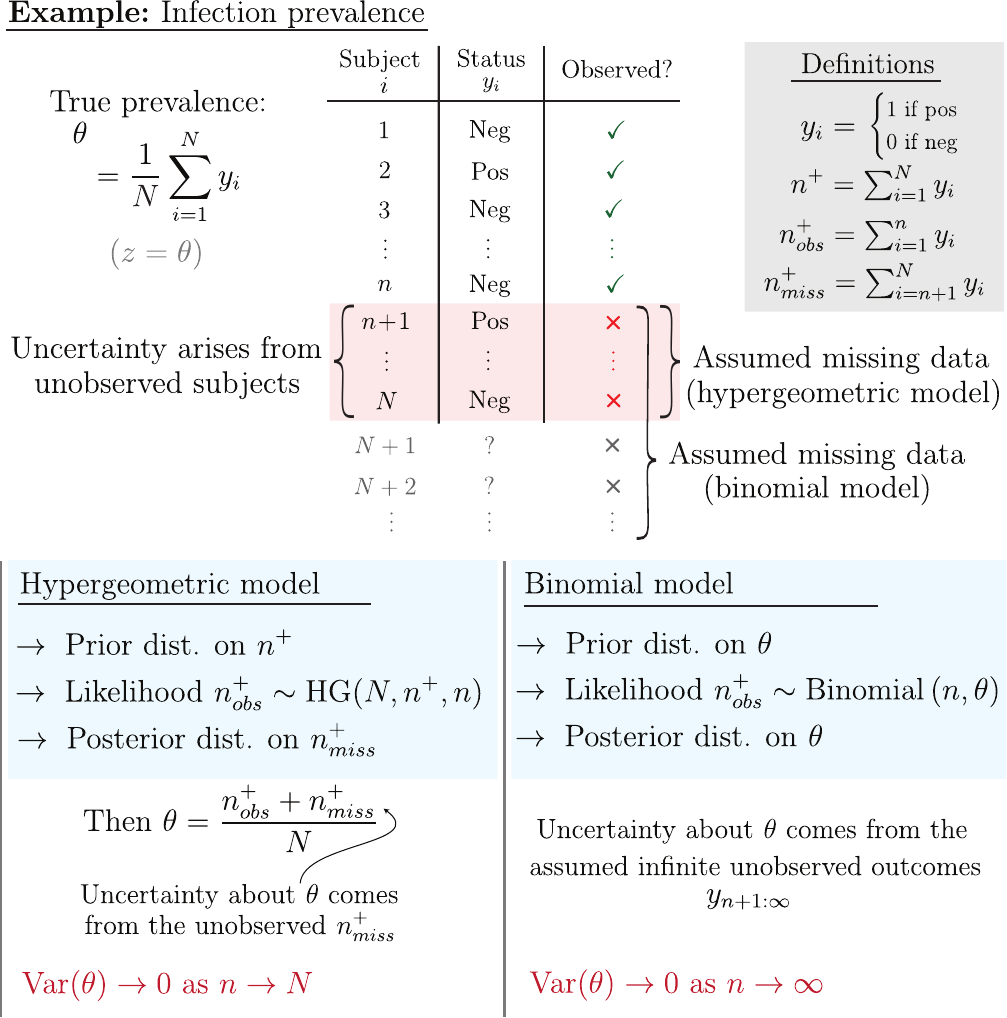}}
    \caption{Uncertainty associated with estimating infection prevalence (here, $z = \theta$). Assume we test $n$ subjects out of a population of size $N$ and observe $n^+_{obs}$ positive results. The missing data are the number of positive results $n_{miss}^+$ in the $N\!-\!n$ subjects that were not tested. A popular approach is to place a prior distribution on $\theta$ and model the observed data using a binomial distribution, resulting in a posterior distribution whose variance goes to $0$ as $n \to \infty$. In this example, the assumed missing data are the infection status of an infinite number of untested individuals. An alternative approach is to place a prior distribution on the true total number of infected subjects $n^+$ and model the observed data using a hypergeometric distribution, directly resulting in a posterior distribution on $n_{miss}^+$. By setting $\theta = \frac{n^+_{obs} + n^+_{miss}}{N}$, we obtain a posterior distribution on $\theta$ whose variance (one possible measure of uncertainty) goes to $0$ as $n \to N$. In this example, the assumed missing data are the infection status of the finite number of untested individuals.}
    \label{fig:example1}
\end{figure}

Consider estimating infection prevalence $\theta$ (Figure \ref{fig:example1}). If infection statuses for the full population were known, there would be no uncertainty about $\theta$. With only a subset tested, a binomial model treats uncertainty as arising from an infinite unobserved population, while a hypergeometric model treats it as directly arising from the finite unobserved population. In both cases, uncertainty flows from assumed missing information, with the precise nature determined by the model.

\subsubsection*{Principle 1: Model-based uncertainty arises from decision-making under incomplete information.}

Decision theory directly relates the missing data to a real-world decision, giving rise to concrete and problem-grounded measurements of uncertainty. For simplicity, we take the decision to be making an estimate (covering both inference and prediction), where the statistical estimator performs the role of a decision rule that maps observed data to some form of estimate. The framework is readily extended to other types of decisions (such as whether to impose an intervention), where the decision rule maps observed data to a real-world action.

In this framework, the modeller's role is to make an estimate $a \in \mathcal{A}$ of unknown $z \in \mathcal{Z}$, inducing a loss $\ell(a, z)$ to be minimised, where $\mathcal{A}$ is the space estimates (actions) that could be made and $\mathcal{Z}$ are the possible values of $z$. $\ell(a,z)$ encodes the real-world consequence of making an estimate $a$ when the true value is $z$, a concept most familiar to machine-learning modellers and statistical forecasters \cite{gneitingStrictlyProperScoring2007, hastieElementsStatisticalLearning2009}. The unknown quantity $z$ could be a model parameter (e.g., $R_t$) or a future observation (e.g., future hospitalisations).

The decision-theoretic approach begins by fitting a model $p(z;y_{obs})$ to the observed data $y_{obs}$. In the prevalence example, this is the posterior distribution $p(\theta|y_{1:n})$. The \textit{Bayes-optimal} estimate $a^*$ minimises the expected loss $E_{p(z|y_{obs})}[\ell(a, z)]$ under the fitted model:
\begin{equation}
    a^* = \arg\min_{a \in \mathcal{A}} E_{p(z|y_{obs})}[\ell(a, z)].
\end{equation}

For point estimates ($\mathcal{A} = \mathcal{Z}$), the quadratic loss $\ell(a, z) = (a - z)^2$ is commonly used, whose Bayes-optimal estimate is the posterior mean. For probabilistic estimates ($\mathcal{A} = \mathcal{P}(\mathcal{Z})$, the set of probability distributions on $\mathcal{Z}$), the log-loss $\ell(a, z) = -\log a(z)$ is commonly used, whose Bayes-optimal estimate is the posterior distribution itself.

A key insight of \textcite{degrootUncertaintyInformationSequential1962} is that the minimised expected loss provides a general and problem-grounded definition of uncertainty \cite{bickfordsmithRethinkingAleatoricEpistemic2025, dawidCoherentMeasuresDiscrepancy1998}. Borrowing notation from \textcite{bickfordsmithRethinkingAleatoricEpistemic2025}, we define uncertainty $h$ as:
\begin{equation}
\begin{aligned}
    h\left[p(z|y_{obs})\right] &\coloneq \min_{a \in \mathcal{A}} E_{p(z|y_{obs})}[\ell(a, z)],\\
    &= E_{p(z|y_{obs})}[\ell(a^*, z)].
\end{aligned}
\end{equation}
Uncertainty is measured by examining the expected loss incurred by taking the optimal action given the information at hand. 

\begin{figure}[H]
    \centering
    \fbox{\includegraphics[width=0.9\columnwidth]{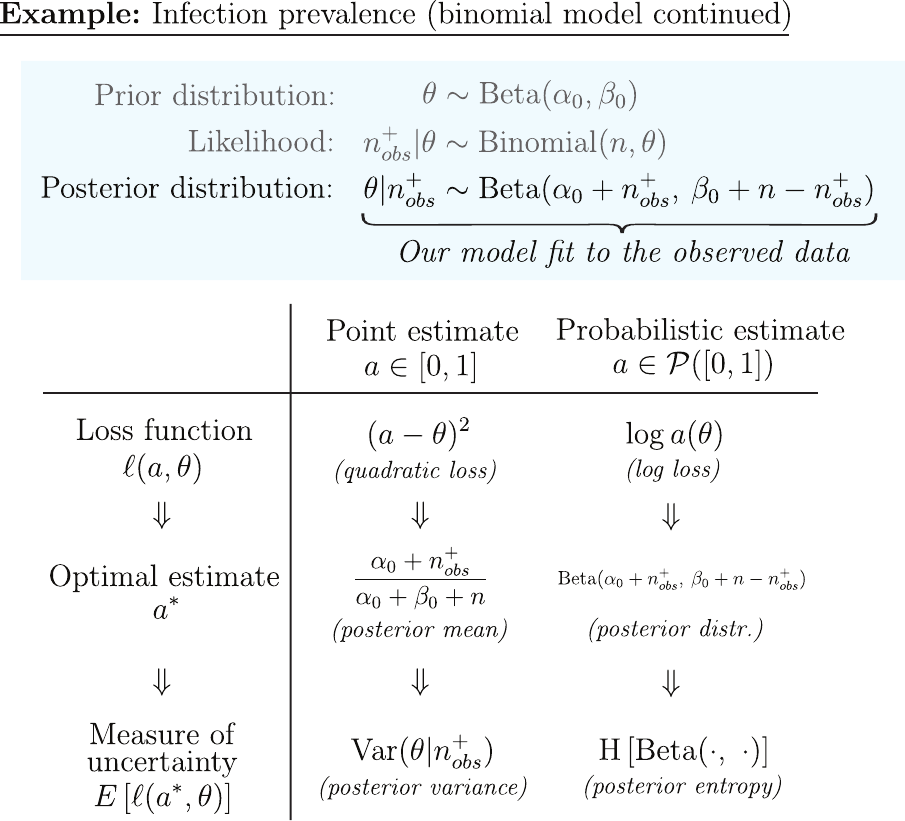}}
    \caption{\textbf{Principle 1}. Two measures of uncertainty in the infection prevalence example with the binomial model ($z = \theta$ is the estimand). For simplicity, we only show the binomial model. If we use a conjugate $\text{Beta}(\alpha_0, \beta_0)$ prior distribution, then the posterior distribution is $\text{Beta}(\alpha_0 + n_{obs}^+, \beta_0 + n - n_{obs}^+)$. If we make a point estimate of $\theta$ under the quadratic loss function, then the Bayes-optimal estimate is the posterior mean and uncertainty can be measured using the posterior variance. If we make a probabilistic estimate of $\theta$ under the log-loss, then the Bayes-optimal estimate is the posterior distribution itself and uncertainty can be measured using the posterior (differential) entropy.}
    \label{fig:example2}
\end{figure}

Multiple popular measures of uncertainty can be derived from this setup \cite{bickfordsmithRethinkingAleatoricEpistemic2025, jacksonValueInformationAnalysis2022, dawidCoherentMeasuresDiscrepancy1998}. For a point estimate and quadratic loss where $a^*$ is the posterior mean, uncertainty is measured by the posterior variance:
\begin{equation}
\begin{aligned}
    h\left[p(z|y_{obs})\right] &= E_{p(z|y_{obs})}[(E_{p(z|y_{obs})}[z] - z)^2],\\
    &= \text{Var}_{p(z|y_{obs})}(z).
\end{aligned}
\end{equation}
For a probabilistic estimate and log-loss where $a^*$ is the posterior distribution, uncertainty is measured by the differential entropy of the posterior distribution:
\begin{equation}
\begin{aligned}
    h\left[p(z|y_{obs})\right] &= E_{p(z|y_{obs})}[-\log p(z|y_{obs})],\\
    &= \text{H}[p(z|y_{obs})]. \label{eq:entropy}
\end{aligned}
\end{equation}
We illustrate this for the prevalence example in Figure \ref{fig:example2}.

The measure of uncertainty is determined by the loss function, thus alternative functions can capture different consequences of estimation error. When estimating $R_t$, for example, authorities may want to prioritise avoiding underestimation to be conservative about epidemic control. Here, we could use a loss function that penalises underestimation more than overestimation, or target tail risk with a quantile (pinball) loss function \cite{gneitingModelDiagnosticsForecast2023}. This yields problem-specific uncertainty measurements, reflecting user-specific risk preferences, that can guide decision-making, data-collection, and model comparison (for example).

\subsubsection*{Principle 2: The incomplete information driving uncertainty is application-specific and determines the decomposition of reducible and irreducible uncertainty.}

In machine learning and other fields, it is common to assume observed data $\{y_i\}_{i=1}^n$ are independent and identically distributed (i.i.d.) from some ``training'' distribution $y_i \sim p_{train}(y)$. This leads to the notion that irreducible (sometimes called ``aleatoric'' \cite{pennIntrinsicRandomnessEpidemic2023}, although we emphasise this term has many definitions in the literature \cite{bickfordsmithRethinkingAleatoricEpistemic2025}) uncertainty is the uncertainty that would remain after fitting the model to the ``complete'', often infinite, training data. In this view, the missing information corresponds to the unobserved data $y_{n+1:\infty}$.

In epidemiological transmission modelling, we only ever observe one realisation of a given epidemic, making the empirical quantification of inherent uncertainty difficult \cite{pennIntrinsicRandomnessEpidemic2023}. Arguing about repeating the epidemic multiple times is of limited practical value, as this offers little insight into the uncertainty we could hope to reduce in practice. Uncertainty instead must be reduced by collecting higher-quality, or a greater quantity of, data during the epidemic \cite{paragQuantifyingInformationNoisy2022}.

We argue that focusing on missing data $y_{miss}$ that \textit{could practically be observed} is more useful. Specifically, we define the uncertainty reduction as the difference between the uncertainty of the model fit to the observed data ($y_{obs}$) and the uncertainty of the model fit to the full data ($y_{all} = y_{obs} \cup y_{miss}$) \cite{bickfordsmithRethinkingAleatoricEpistemic2025}:

\begin{equation}
    \resizebox{0.9\columnwidth}{!}{$
    \underbrace{\text{UR}_z(y_{miss})}_{\text{Uncertainty reduction}} = \underbrace{h\left[p(z|y_{obs})\right]}_{\text{Total uncertainty}} - \underbrace{h\left[p(z|y_{all})\right].}_{\text{Irreducible uncertainty}} $} \label{eq:UR}
\end{equation}

\begin{figure}[H]
    \centering
    \fbox{\includegraphics[width=\columnwidth]{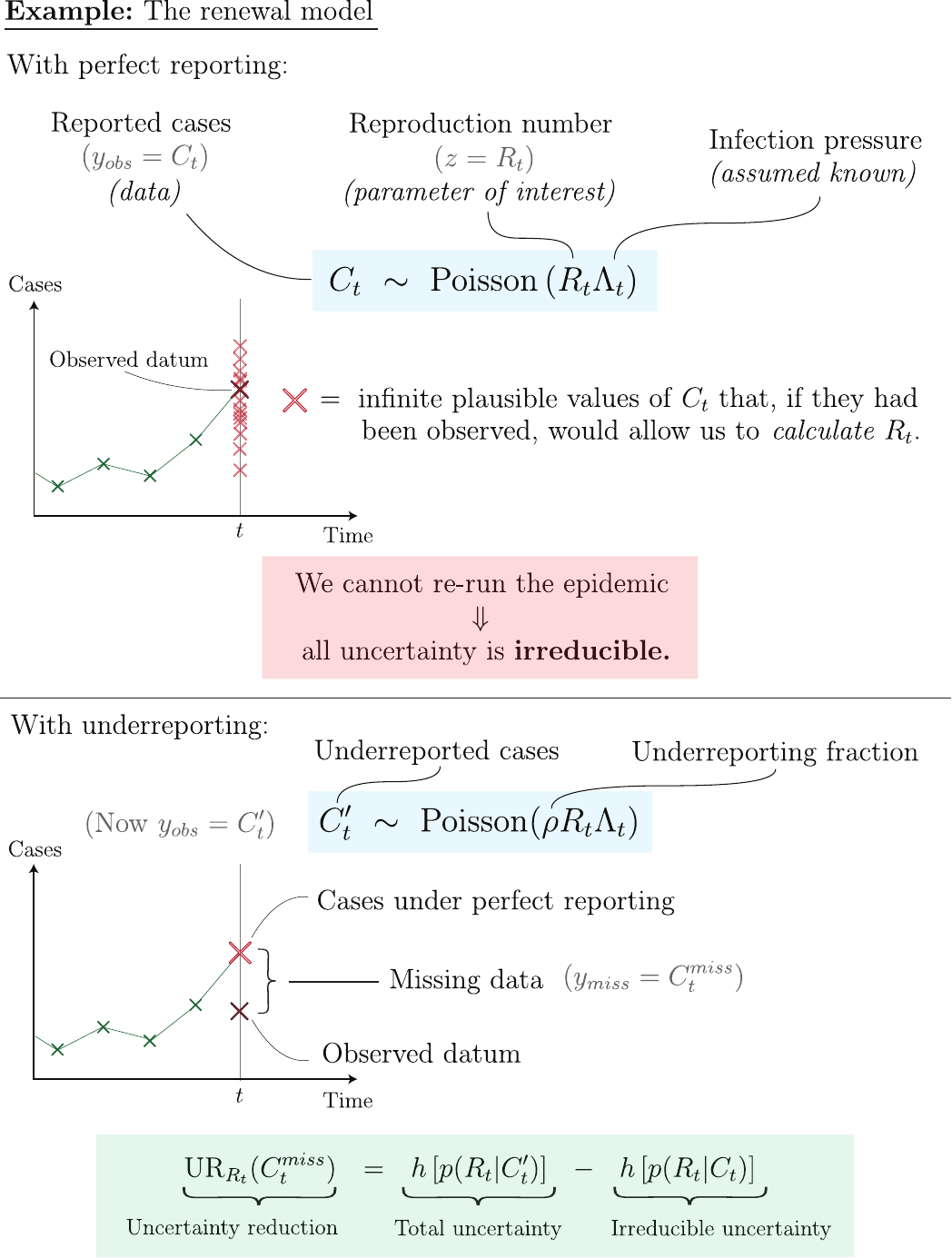}}
    \caption{\textbf{Principle 2.} Reducible and irreducible uncertainty in the renewal model. The instantaneous reproduction number ($R_t$) is a key quantity in epidemiological modelling, defined as the average number of secondary infections caused by an infected individual at time $t$ if they were to undergo their entire infectious period at this time \cite{paragAreEpidemicGrowth2022}. It is commonly estimated from reported case data using the Poisson renewal model pictured here \cite{coriNewFrameworkSoftware2013, paragImprovedEstimationTimevarying2021}. Even if the reported cases $C_t$ perfectly corresponded to the true infections (i.e., no underreporting), uncertainty about $z = R_t$ would still exist. Under this model, uncertainty about $R_t$ could be eliminated only if we were to repeat the epidemic infinitely many times, thus all uncertainty is irreducible. In the presence of underreporting, the missing data are the unreported infections. If we knew these quantities, for example, by improving outbreak surveillance, we could reduce our uncertainty about $R_t$.}
    \label{fig:example3}
\end{figure}

For example, consider estimating $R_t$ using the Poisson renewal model (Figure \ref{fig:example3}) \cite{coriNewFrameworkSoftware2013}. Under perfect case reporting, $R_t$ uncertainty could be eliminated only by repeating the epidemic infinitely many times. Here, all uncertainty is irreducible. In the presence of underreporting, for example, the missing data are unreported cases, and if we knew these, we could reduce $R_t$ uncertainty. Comparing this to the prevalence example, where $y_i$ are assumed to be i.i.d., we see a key difference between epidemiology (and weather forecasting, economics, etc \cite{moranEpidemicForecastingMessier2016, millsMetricActionEvaluation2025}) and other fields such as machine learning: we often only observe a single realisation of the data-generating process and arguing about alternative realisations is of limited practical value.

This construction is purposefully broad. The forms of $p(z|y_{obs})$ and $p(z|y_{all})$ are not specified, allowing any model to be used. We also note that the form of the missing data $y_{miss}$ need not match $y_{obs}$, enabling the inclusion of different datasets in the model. We illustrate this in section \ref{sec:applications}, considering the inclusion of wastewater data in a model fit only to reported cases.

\subsubsection*{Principle 3: To quantify reducible and irreducible uncertainty, we reason about the missing data.}

In some scenarios, we may know the value(s) of $y_{miss}$, for example, when including an additional already collected dataset in a model. In many scenarios, however, such as when planning future data collection, or estimating the irreducible uncertainty, we do not know the value(s) of $y_{miss}$ in advance. In these cases, we must reason about the missing data that could be observed, and consider the \textit{expected} uncertainty reduction.

Letting $p(y_{miss}|y_{obs})$ be a predictive model describing our beliefs about the missing data, we average over all possible realisations of $y_{miss}$ by taking expectations of both sides of Equation \ref{eq:UR}, leading to the expected uncertainty reduction:

\begin{multline}
    \underbrace{\text{E}_{p(y_{miss}|y_{obs})}\left[\text{UR}_z(y_{miss})\right]}_{\text{Expected uncertainty reduction}} =\\
    \underbrace{h\left[p(z|y_{obs})\right]}_{\text{Total uncertainty}} - \underbrace{\text{E}_{p(y_{miss}|y_{obs})}\left[h\left[p(z|y_{all})\right]\right].}_{\text{Expected uncertainty after collecting $y_{miss}$}} \label{eq:EUR}
\end{multline}

\begin{figure}[H]
    \centering
    \fbox{\includegraphics[width=\columnwidth]{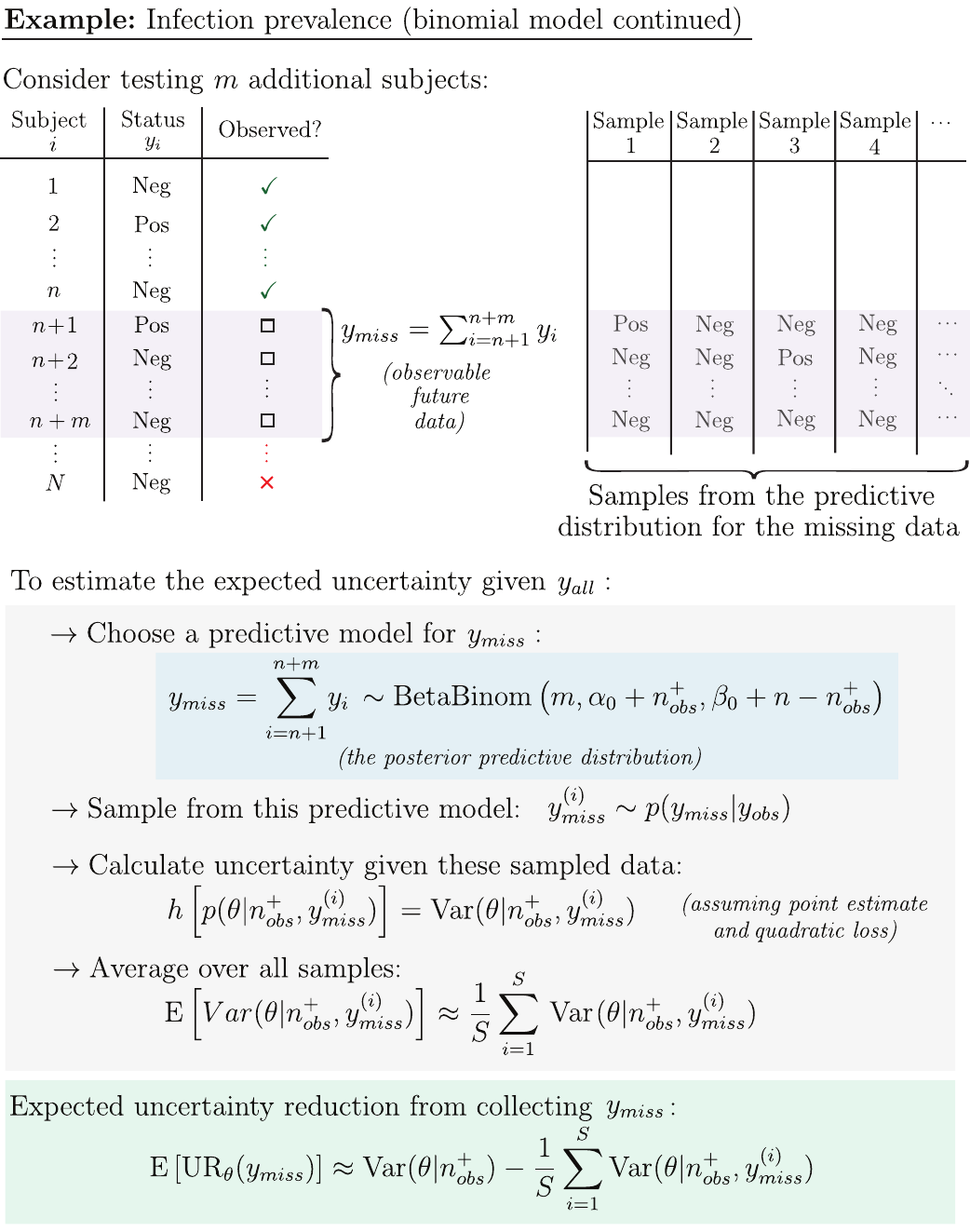}}
    \caption{\textbf{Principle 3.} Estimating the expected uncertainty reduction in the infection prevalence example with the binomial model ($z = \theta$). Consider testing $m$ additional individuals for infection. The posterior-predictive distribution given a beta prior distribution and binomial likelihood is a beta-binomial distribution. We calculate the expected uncertainty reduction about $\theta$ after collecting $y_{miss}$ by averaging $h\left[p(\theta|y_{1:n}, y_{miss})\right]$ (the final term in equation \ref{eq:EUR}) over this predictive distribution.}
    \label{fig:example4}
\end{figure}

In some scenarios, the appropriate predictive distribution may be obvious. For example, if we are constructing a Bayesian model using the standard prior-likelihood-posterior update and are considering the uncertainty reduction associated with collecting the same type of data, the posterior-predictive distribution is appropriate. We illustrate this in the context of prevalence estimation in Figure \ref{fig:example4}. In other scenarios, this predictive distribution may be less clear. In either case, $p(y_{miss}|y_{obs})$ is necessarily subjective and describes our beliefs about the missing data.

\subsubsection*{Principle 4: Collecting additional data is not guaranteed to reduce uncertainty in practice, but will in expectation if the predictive distribution is \textit{coherent}.}

In practice, new data may contradict existing data, leading to an increase in uncertainty about $z$. While this cannot be prevented with certainty, if the predictive distribution $p(y_{miss}|y_{obs})$ is \textit{coherent} with the fitted model, then additional data will reduce uncertainty \textit{in expectation}. 

$p(y_{miss}|y_{obs})$ is coherent with the fitted model $p(z|y_{obs})$ if both are marginal distributions of the same joint distribution $p(z, y_{all})$, implying:
\begin{equation}
    p(z|y_{obs}) = \int_\mathcal{Y} p(z|y_{all}) p(y_{miss}|y_{obs}) \ dy_{miss}, \label{eq:coherencedefn}
\end{equation}
where $\mathcal{Y}$ is the space of possible observed $y_{miss}$.

This means that the predictive distribution and the fitted model make the same assumptions about the missing data, and that the expected value of $z$ should not change after including missing data sampled from $p(y_{miss}|y_{obs})$.

To see that coherence leads to an expected uncertainty reduction, define $a^*$ as the Bayes-optimal estimate of $z$ prior to observing $y_{miss}$ (but after observing $y_{obs}$). After observing $y_{miss}$, the Bayes-optimal estimate may change, giving the first inequality:

\begin{equation}
\resizebox{\columnwidth}{!}{$
\begin{aligned}
    \min_a \text{E}_{p(z|y_{all})}\left[\ell(a,z)\right] &\leq \text{E}_{p(z|y_{all})}\left[\ell(a^*,z)\right]\\
    \implies \text{E}_{p(y_{miss}|y_{obs})}\left\{\min_a \text{E}\left[\ell(a,z)\right]\right\} &\leq \text{E}_{p(y_{miss}|y_{obs})}\left\{\text{E}\left[\ell(a^*,z)\right]\right\}\\
    &= \text{E}_{p(z|y_{obs})}\left[\ell(a^*,z)\right].
\end{aligned}
$}
\end{equation}

Taking expectations over $p(y_{miss}|y_{obs})$ gives the second inequality, and the final equality follows from coherence and the law of total expectation. Re-writing this in terms of $h$ gives $\text{E}\left[h(p(z|y_{all}))\right] \leq h(p(z|y_{obs}))$, thus the expected uncertainty reduction is non-negative. Equality holds if and only if $z$ is conditionally independent of $y_{miss}$ given $y_{obs}$. That is, the expected uncertainty reduction is zero only when the missing data provide no new information about $z$.

This condition re-asserts the missing data as the source of uncertainty. While we can use any model of the missing data $p(y_{miss}|y_{obs})$ to quantify the expected uncertainty reduction, we may get counterintuitive results if our predictive model is not coherent with the fitted model. Intuitively, coherence ensures that the explicit assumptions about the missing data that we make when separating uncertainty are consistent with the implicit assumptions made by the fitted model.

In the prevalence example, suppose we believe the prevalence in individuals who do not volunteer for testing may be different from those who do. Reflecting this, we decide to allow for additional variation in the missing data by inflating the variance of the posterior-predictive distribution. If this increase in variance is introduced only in the prediction step, but not in the fitted model, then the predictive model is not coherent with the fitted model, and uncertainty may increase on average.

Coherence is a fundamental property of Bayesian inference and is often used to argue for Bayesian approaches over other inferential approaches. In fact, a model satisfies the coherence condition if and only if it is Bayesian \cite{robinsConditioningLikelihoodCoherence2000, bissiriGeneralFrameworkUpdating2016} (i.e., it updates beliefs via Bayes' rule). That is, the expected uncertainty reduction defined by Equation \ref{eq:EUR} is non-negative if and only if our model is Bayesian. While this makes using standard Bayesian approaches appealing, this also allows us to use any model that satisfies coherence (Equation \ref{eq:coherencedefn}) for Bayesian uncertainty quantification \cite{shirvaikarPredictionpoweredMachineLearning2025}.

\begin{figure}[h!]
    \centering
    \fbox{\includegraphics[width=0.75\textwidth]{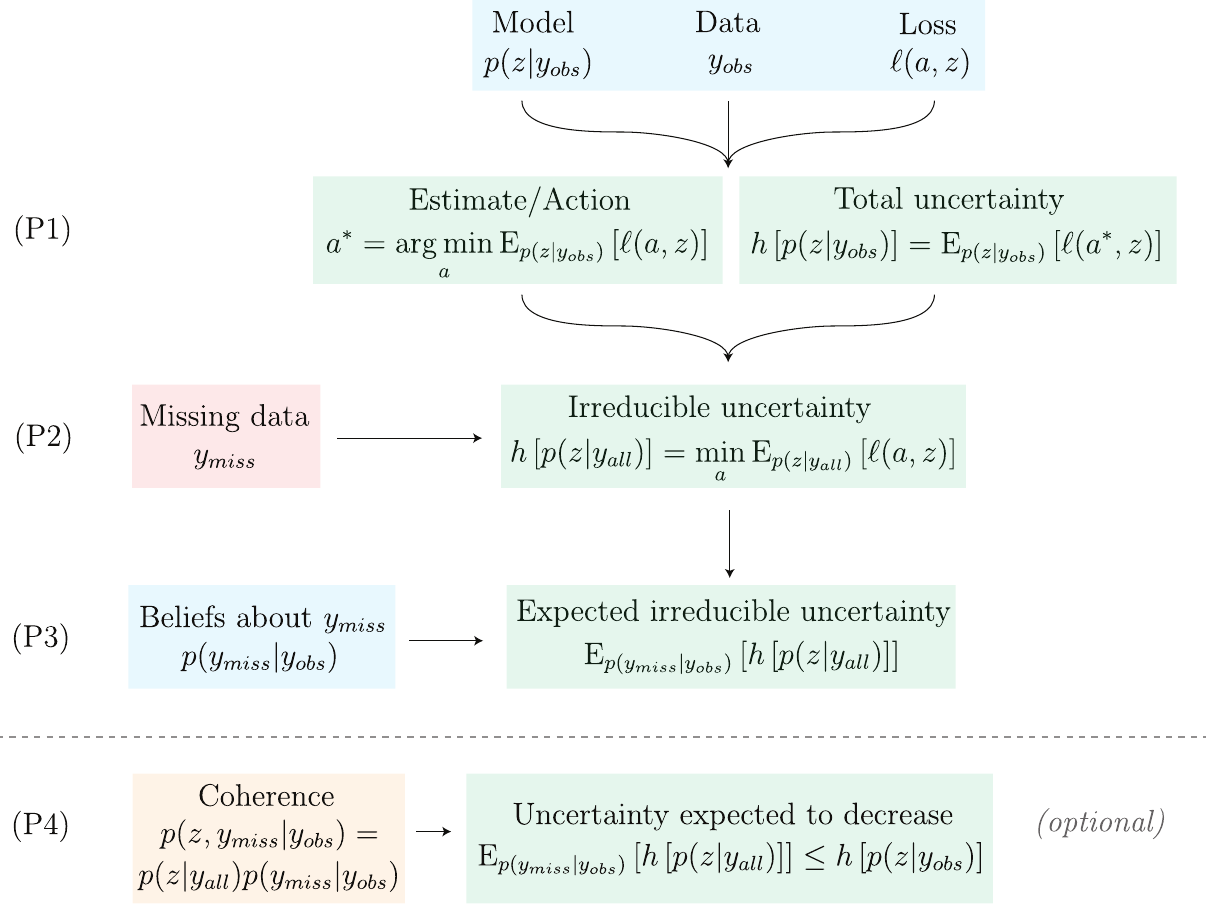}}
    \caption{An overview of the four principles (P1 to P4) of our decision-theoretic framework. We start with a model $p(z|y_{obs})$ of the unknown quantity $z$, observed data $y_{obs}$, and a loss function $\ell(a, z)$ - the three quantities needed to make a decision under uncertainty. P1 uses the loss function to concretely relate the model and data to the real-world act of making an estimate $a$ of $z$, defining the Bayes-optimal estimate $a^*$ and the corresponding total uncertainty $h$. P2 defines the irreducible uncertainty as the uncertainty remaining after collecting missing data $y_{miss}$. As we generally do not know the value of $y_{miss}$, P3 introduces the expected irreducible uncertainty, which is the irreducible uncertainty averaged over a predictive distribution $p(y_{miss}|y_{obs})$ of the missing data. P4 provides an optional modelling condition, related to the formulation of $p(y_{miss}|y_{obs})$, under which collecting these missing data is expected to reduce uncertainty. Blue shading highlights known inputs, red shading highlights possibly unknown inputs, green shading highlights framework outputs, and yellow shading highlights the optional coherence condition.}
    \label{fig:diagram1}
\end{figure}

We emphasise that coherence is not a requirement of our framework and principles (1) to (3) are valid regardless. However, without this assumption, collecting more data is not necessarily expected to reduce uncertainty. All four principles are summarised in Figure \ref{fig:diagram1}.

\end{multicols}

\clearpage

\begin{multicols}{2}

\section{An application to wastewater surveillance}
\label{sec:applications}

\textbf{How much can wastewater surveillance reduce uncertainty about $R_t$?}

Wastewater surveillance gained prominence during the COVID-19 pandemic as a potentially less biased source of data than traditional sources (such as reported cases) for tracking real-time pathogen transmission \cite{millsUtilityWastewaterSurveillance2024, watsonJointlyEstimatingEpidemiological2024}. The concentration of genomic material in wastewater does not map directly onto traditional epidemiological indicators and can be highly variable \cite{bertelsFactorsInfluencingSARSCoV22022}. This has led some to question the utility of wastewater data in this setting \cite{millsUtilityWastewaterSurveillance2024}.

\textcite{watsonJointlyEstimatingEpidemiological2024} constructed a joint model of wastewater data (collected by the New Zealand Institute for Public Health and Forensic Science \cite{hewittSensitivityWastewaterbasedEpidemiology2022, theinstituteofenvironmentalscienceandresearchCOVID19DataRepository}) and reported case data for Aotearoa New Zealand \cite{nzministryofhealthCOVID19DataNew2025}. To examine the utility of wastewater data for estimating $R_t$, we apply our framework to answer two questions: first, to what extent is uncertainty about $R_t$ reduced by including wastewater data in the model (compared to modelling reported cases only)? And second, given current wastewater sampling in Aotearoa New Zealand, how much could uncertainty about $R_t$ be reduced by sampling wastewater catchments covering the entire population every day? We assume decision-makers use point estimates of $R_t$, with a quadratic loss function deemed suitable, and therefore use variance as our measure of uncertainty.

To assess the benefit of incorporating already-observed data, we compute the daily uncertainty reduction (Equation \ref{eq:UR}). In this setting, $y_{miss}$ are the daily waste\-water observations $W_{1:T}$, and $y_{obs}$ are daily reported case data $C_{1:T}$. The uncertainty reduction in $R_t$ at time-step $t \in 1, 2, \ldots T$ is given by $\text{Var}_{p(R_t|C_{1:T})}(R_t) - \text{Var}_{p(R_t|C_{1:T}, W_{1:T})}(R_t)$, where $p(R_t|C_{1:T})$ is the posterior distribution of $R_t$ given only the reported case data, and $p(R_t|C_{1:T}, W_{1:T})$ is the posterior distribution of $R_t$ given both the reported case data and the wastewater data.

To estimate the expected uncertainty reduction in $R_t$ from daily wastewater sampling of the entire population, we first fit the model to the observed data $C_{1:T}$ and $W_{1:T}$, obtaining the joint posterior distribution over infection incidence $I_{1:T}$ and nuisance parameter vector $\theta$, denoted $p(I_{1:T}, \theta|C_{1:T}, W_{1:T})$. We then simulate hypothetical missing wastewater data $W_{1:T}'$ by assuming the daily catchment population equals the missing population and sampling from the posterior predictive distribution $p(W_{1:T}'|C_{1:T}, W_{1:T}) = \int \int P(W_{1:T}'|I_{1:T}, \theta) P(I_{1:T}, \theta|C_{1:T}) dI_{1:T} d\theta$. We fit the model to the simulated dataset (reported cases and the daily population-weighted average of $W_{1:T}$ and $W_{1:T}'$), calculate the associated uncertainty, and repeat this 100 times to obtain a Monte Carlo estimate of the expected uncertainty reduction.

\begin{figure}[h!]
    \centering
    \includegraphics[width=\textwidth]{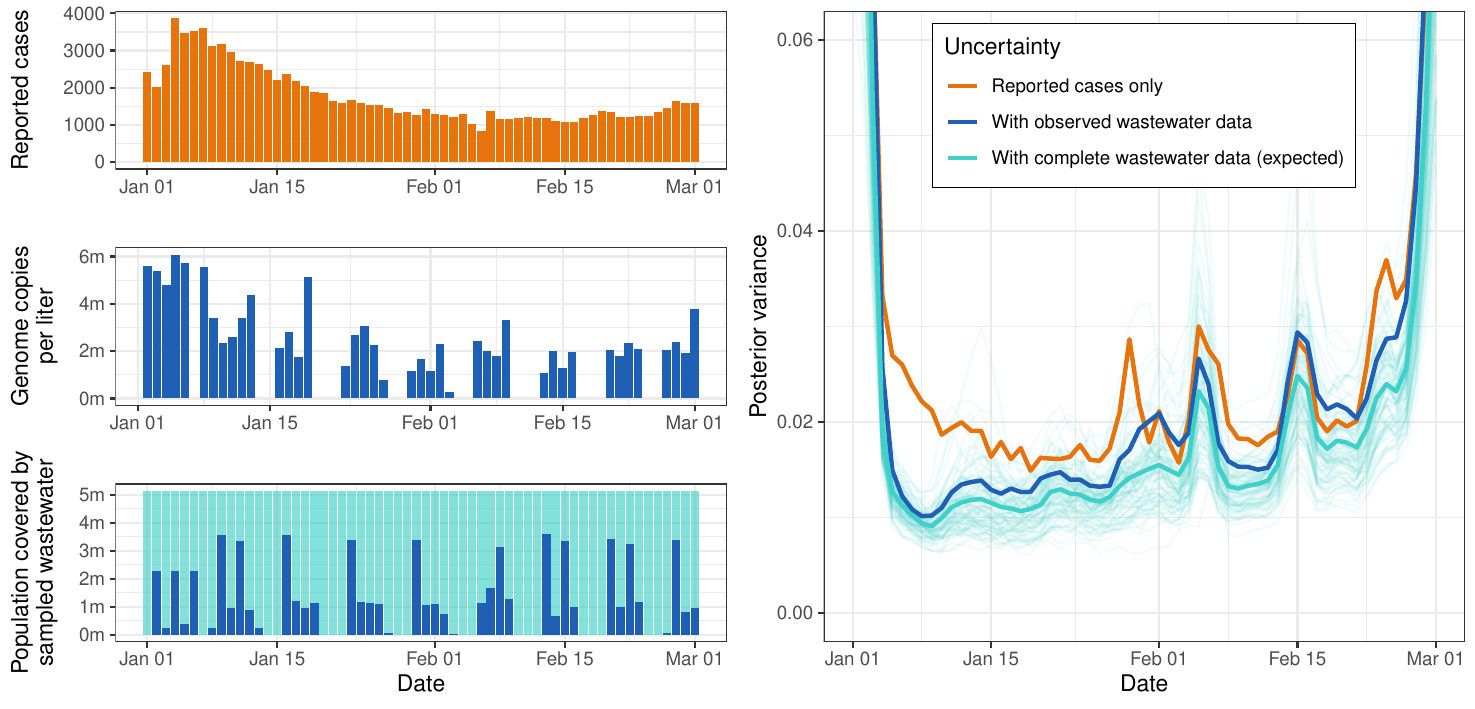}
    \caption{Reported case data (A), wastewater data (B), and population covered by wastewater catchment areas (C) for Aotearoa New Zealand between 1 January 2023 and 1 March 2023. The light blue region in panel (C) represents the population from which missing data could be sampled. The daily uncertainty about $R_t$ from the model fit to reported case data only (orange), from the model fit to both reported case data and observed wastewater data (dark blue), and the expected uncertainty after sampling wastewater data from the entire population every day (light blue) are shown in panel (D). Thin light blue lines show the posterior variance of $R_t$ from individual simulations of hypothetical data, demonstrating that even though the expected uncertainty reduction is positive (as we satisfy the coherence requirements), uncertainty about $R_t$ could still plausibly increase. Uncertainty increases at the start and end of the time period as a result of fewer data points being available to fit the model.}
    \label{fig:wastewater}
\end{figure}

We fit the model to data between 1 January 2023 and 1 March 2023 (Figure \ref{fig:wastewater}-A), sourced from \cite{watsonJointlyEstimatingEpidemiological2024}. Wastewater data were collected on 41 out of these 60 days (Figure \ref{fig:wastewater}-B), with daily catchment coverage ranging from 31,000 to 3.6 million individuals (out of 5.15 million total, Figure \ref{fig:wastewater}-C).

Including wastewater data reduces expected daily uncertainty about $R_t$ by an average 16.9\%, with the greatest reduction being 54.4\% on 7 January 2023, though some increases (up to 12.6\% on 31 January 2023, 12.1\% on 17 and 18 February 2023) occur on some days (Figure \ref{fig:wastewater}-D). Sampling wastewater data from the entire population every day is expected to further reduce daily uncertainty by 14.6\% (or 29.4\% when compared to estimates from reported case data only), representing a plausible bound on the expected reduction in $R_t$ uncertainty achievable through wastewater sampling.

Missing wastewater data may (once collected) differ systematically from observed data, for example if existing sampling is biased towards sites expected to yield more reliable measurements. In this case, we might wish to allow the variance of the observation distribution for the wastewater data to be larger in the hypothetical-data setting than in the observed data. To maintain coherence and ensure more data reduces our model uncertainty, we would also need to include this assumption in the model itself.

\clearpage

\end{multicols}

\clearpage

\begin{multicols}{2}

\section{Related frameworks}

Our decision-theoretic approach captures multiple popular methodologies in epidemiology under a single umbrella. We outline some of these here.

The \textbf{value of information} (VOI) framework, widely used in health economics, measures the expected reduction in loss from obtaining additional data \cite{jacksonValueInformationSensitivity2019, jacksonValueInformationAnalysis2022, heathValueInformationHealthcare2024, jacksonGuideValueInformation2021, fenwickValueInformationAnalysis2020, raiffaAppliedStatisticalDecision1961}. A central concept in the VOI framework is the \textit{expected value of sample information} (EVSI), which quantifies the expected reduction in loss from collecting new data $y_{new}$, written in our notation as:
\begin{equation}
    \resizebox{0.9\columnwidth}{!}{$
    \text{EVSI}(y_{new}) = h\left[p(z|y_{obs})\right] - \text{E}_{p(y_{new}|y_{obs})}\left[h\left[p(z|y_{obs}, y_{new})\right]\right]
    $}
\end{equation}
The expectation is taken with respect to the posterior predictive distribution for the new data $p(y_{new}|y_{obs})$, thus the EVSI is equivalent to our Equation \ref{eq:EUR} in a standard Bayesian setting. Our framework extends the VOI literature beyond the traditional Bayesian setting by allowing for a wider class of predictive distributions and models to be used for valid uncertainty quantification.

The VOI framework is also applied to decision problems, where the loss function is defined in terms of the cost of making a decision based on the model. Our framework adds a layer of interpretation to this setting: the minimised expected cost of the decision is a direct measure of uncertainty about the decision being made.

\textbf{Bayesian experimental design} and \textbf{active learning} are focused on data collection strategies, arguing that experiments should be designed to maximise their expected utility, commonly framed in terms of the \textit{expected information gain} (EIG) \cite{rainforthModernBayesianExperimental2024}. Letting $\xi$ denote the experimental design that collects data $y_{new}$ (a yet-to-be-observed random variable), the EIG about $z$ is defined (in our notation) as:
\begin{equation}
\resizebox{0.9\columnwidth}{!}{$
\begin{aligned}
    \text{EIG}_z(\xi) &\coloneq \text{E}_{p(y_{new}\mid \xi)} \left[ \log p(z\mid y_{new}, y_{obs}, \xi) - \log p(z\mid y_{obs}) \right] \\
    &= \text{E}_{p(y_{new}\mid \xi)} \left[ \text{H} \left[p(z\mid y_{obs})\right] - \text{H} \left[p(z\mid y_{new}, y_{obs}, \xi)\right] \right],
\end{aligned}
$}
\end{equation}

where $\text{H} \left[p(z\mid y_{obs})\right] - \text{H} \left[p(z\mid y_{new}, y_{obs}, \xi)\right]$ is the information gain about $z$ from observing $y_{new}$ from experiment $\xi$. This can be viewed as a special case of Equation \ref{eq:EUR} under a log-loss function. By Principle 4, an experiment $\xi$ is only guaranteed to reduce uncertainty if it is equivalent to sampling from the posterior predictive distribution, opening a discussion on model specification which we leave for future work. Adaptive learning extends the method to allow for updating $\xi$ as data are collected.

\textcite{caseAdaptingVectorSurveillance2024} apply experimental design principles in epidemiology, specifically in designing tick surveillance strategies, although they do not directly target the EIG. Instead, they target two criteria: Bayesian d-optimality (approximately equivalent to maximising the EIG \cite{alexanderianBriefNoteBayesian2023}) and a bespoke loss function that minimises the maximum standard deviation of estimates in ``high-risk'' regions. Our framework encourages viewing the expected value of their bespoke loss function as the relevant measure of uncertainty in this setting.

\textbf{Fisher information} approaches can also be captured by our framework. For example, \textcite{paragQuantifyingInformationNoisy2022} use information-theoretic methods to quantify the information in noisy epidemic curves, which they define as the Fisher information (FI) of $R_t$. The FI is approximately equivalent to the inverse-variance of the maximum likelihood estimate, so this approach effectively measures uncertainty in terms of the variance of a point estimate. The equivalent approach under our framework is the use of a quadratic loss function, examining the expected uncertainty (in this example, variance) reduction about $R_t$ after reducing reporting delays and noise.

Finally, \textbf{scoring rules} have gained popularity in epidemiology for measuring the quality and value of probabilistic estimates \cite{bosseEvaluatingForecastsScoringutils2024, millsMetricActionEvaluation2025, gneitingStrictlyProperScoring2007, banholzerComparisonShorttermProbabilistic2023, steynRobustUncertaintyQuantification2025}. Estimation of future data is distinct from estimation of model parameters in that we eventually observe the data, and thus we can eventually \textit{calculate} the loss (score) associated with a given prediction. In our framework, applying a scoring rule to each forecast-outcome pair can be viewed as an operationalisation of the loss, thus giving a setting in which we can empirically estimate the uncertainty.

\section{Discussion}

We have presented a decision-theoretic framework for conceptualising and quantifying uncertainty in epidemiological models. Starting from the idea that uncertainty arises from missing data \cite{fongMartingalePosteriorDistributions2023}, our approach directly links theory to practical application and yields a separation of reducible and irreducible uncertainty. We then demonstrated that coherence ensures collecting additional data will reduce uncertainty on average, reinforcing missing data as the source of uncertainty. The framework was illustrated using a published model of SARS-CoV-2 wastewater surveillance in Aotearoa New Zealand \cite{watsonJointlyEstimatingEpidemiological2024}, where the practical limit of wastewater data in reducing $R_t$ uncertainty was estimated.

By generalising, extending, and improving existing frameworks in epidemiology, we broaden their applicability, clarifying how divergent perspectives on uncertainty often overlap. For example, while VOI is usually applied to traditional Bayesian models, our results allow a wider range of models to be used, and Principle 4 provides guidance on when the EVSI is expected to be non-negative. The VOI literature also introduces concepts such as the \textit{expected value of perfect information} (EVPI), which quantifies the loss reduction if the model parameters were known exactly. This also induces a reducible/irreducible split, though we argue this split is less useful in our context than focusing on reductions available through data collection.

This work lays the foundation for potentially high-impact future research, including (i) application-specific loss functions, (ii) general coherent predictive models, and (iii) the evaluation of model specification. For (i), adopting losses that reflect the decision context, or even the decision itself (instead of an estimate), could lead to more relevant notions of uncertainty. In particular, there are many cases in epidemiology where being above a certain threshold is of greater concern than being below it \cite{beregiOptimalAlgorithmsControlling2025}. For (ii), while we focused on Bayesian examples, the framework supports a much broader class of models, including black-box machine-learning models. Ongoing work on martingale predictive distributions and conditionally identically distributed sequences is building our understanding of the precise coherence properties necessary to support valid uncertainty quantification, a promising line of research for epidemiological applications \cite{fongMartingalePosteriorDistributions2023,fortiniExchangeabilityPredictionPredictive2025,shirvaikarPredictionpoweredMachineLearning2025, battistonNewOldPredictive2025}. For (iii), as epidemiological models are typically fit to a single data realisation, model structure necessarily plays a greater role in producing estimates than in many other fields. There is consequently greater potential for model misspecification in epidemiology, addressable by extending our framework. Potential research directions include the use of reference distributions \cite{bickfordsmithRethinkingAleatoricEpistemic2025}, parametric extensions \cite{jacksonFrameworkAddressingStructural2011}, and model averaging \cite{jacksonFrameworkAddressingStructural2011, hoetingBayesianModelAveraging1999}. 

In many epidemiological applications, the predictive distribution for $y_{miss}$ will take the form of a simulator. In this setting, the expected uncertainty reduction can be estimated by sampling from this simulator, re-fitting the model, calculating the uncertainty, and repeating this multiple times. Averaging the uncertainties gives a Monte Carlo estimate of the expected uncertainty after collecting $y_{miss}$. This can be computationally expensive, particularly if the simulator is complex. Existing techniques, such as Gaussian process surrogates or neural density estimators \cite{kennedyBayesianCalibrationComputer2001, swallowChallengesEstimationUncertainty2022,papamakariosNeuralDensityEstimation2019}, could accelerate this process.

By framing uncertainty as a consequence of estimation under incomplete information, we provide a theoretically grounded and practical approach to uncertainty quantification in epidemiological modelling, particularly useful for planning studies and directing surveillance efforts. Our framework supports consistent, reproducible, and policy-relevant model-based inference in infectious disease epidemiology.

\section*{Supporting information}

All code and data to reproduce the wastewater application are available at: \url{https://github.com/nicsteyn2/DecisionTheoreticEpi}. Wastewater data were collected by the New Zealand Institute for Public Health and Forensic Science wastewater COVID-19 surveillance programme: \url{https://github.com/ESR-NZ/covid_in_wastewater}. Reported cases are available from the New Zealand Ministry of Health: \url{https://github.com/minhealthnz/nz-covid-data}.

\section*{Acknowledgements}

N.S. acknowledges support from the Oxford-Radcliffe Scholarship from University College, Oxford, the EPSRC CDT in Modern Statistics and Statistical Machine Learning (Imperial College London and University of Oxford) (EP/S023151/1), and A. Maslov for studentship support. F.B.S. acknowledges support from the EPSRC CDT in Autonomous Intelligent Machines and Systems (EP/L015897/1). C.M was supported by a studentship (EP/T517811/1) from the UK EPSRC. V.S. is supported by the EPSRC CDT in Modern Statistics and Statistical Machine Learning (EP/S023151/1) and Novo Nordisk. C.A.D. acknowledges support from the NIHR Health Protection Research Unit in Emerging and Zoonotic Infections and the Oxford Martin Programme in Digital Pandemic Preparedness. K.V.P. acknowledges funding from the MRC Centre for Global Infectious Disease Analysis (Reference No. MR/X020258/1) funded by the UK Medical Research Council. This UK-funded grant is carried out in the frame of the Global Health EDCTP3 Joint Undertaking. The funders had no role in study design, data collection and analysis, decision to publish, or manuscript preparation.

\printbibliography

\end{multicols}

\end{document}